\documentclass[twocolumn,prb,amsmath,amssymb,showpacs]{revtex4}

\usepackage{bm}
\usepackage{graphicx}

\def \be{\begin{equation}}
\def \ee{\end{equation}}
\def \bmlett{\begin{subequations}}
\def \emlett{\end{subequations}}

\def \ve{\varepsilon}

\def \NN{{\mathcal N}}

\def \CC{{\mathcal C}}

\def \psitild{\widetilde{\psi}}

\def \pd{\phantom{\dagger}}

\def \ua{\uparrow}
\def \da{\downarrow}
\def \ra{\rightarrow}

\begin{document}



\title{Quantum-Limited Measurement and Information in Mesoscopic Detectors}

\author{A. A. Clerk, S. M. Girvin and A. D. Stone}
\affiliation{ Departments of Applied Physics and Physics, Yale
University, New Haven CT, 06511, USA\\ Oct. 30, 2002}

\begin{abstract}
We formulate general conditions necessary for a linear-response
detector to reach the quantum limit of measurement efficiency,
where the measurement-induced dephasing rate takes on its minimum
possible value.  These conditions are applicable to both
non-interacting and interacting systems.  We assess the status of
these requirements in an arbitrary non-interacting scattering
based detector, identifying the symmetries of the scattering
matrix needed to reach the quantum limit.  We show that these
conditions are necessary to prevent the existence of information
in the detector which is not extracted in the measurement process.
\end{abstract}

 \maketitle

\section{Introduction}
Issues of quantum measurement in mesoscopic systems have recently
garnered considerable interest, both because of their relevance to
attempts at quantum computation \cite{MakhlinRMP} and
quantum-limited amplifiers \cite{DevoretNature}.  A general
consequence of any quantum measurement is that it must induce
decoherence in the system variable conjugate to that being
measured.  This basic fact naturally leads to the issue of
measurement efficiency: what conditions must a particular detector
satisfy so that it induces the {\it absolute minimum} amount of
dephasing required by quantum mechanics?  This minimum dephasing
rate is identical to the measurement rate $\Gamma_{\rm meas}$, the
rate at which information is extracted during the measurement
process; thus, the measurement efficiency ratio $\chi \leq 1$ is
defined by $\chi = \Gamma_{\rm meas} / \Gamma_{\varphi}$, where
$\Gamma_{\varphi}$ is the measurement-induced dephasing rate.
Besides being of great conceptual interest, near-ideal measurement
schemes are necessary to detect signatures of coherent qubit
oscillations in the output noise of a detector
\cite{Averin,Korotkov}, and are essential if one wishes to
construct a quantum limited amplifier (i.e.~an amplifier whose
noise energy is the minimum allowed by quantum mechanics)
\cite{DevoretNature}. While the question of measurement efficiency
has received attention in the context of general measurement
theory \cite{Braginsky}, it is only recently that it has been
considered in the context of solid state detectors. Averin
\cite{Averin} has considered the status of the quantum limit in a
number of solid state detectors, while recently Pilgram and
B\"{u}ttiker \cite{Pilgram} considered the quantum limit for a
system in which a mesoscopic conductor acts as a detector.

In this paper, we formulate general conditions which are needed
for an arbitrary detector in the linear-response regime to reach
the quantum limit of detection, where $\chi = 1$.  These general
conditions are valid for both interacting and non-interacting
systems, and can be given a direct physical interpretation.  We
also discuss the quantum limit in terms of a simple concept from
quantum information theory, the accessible information.  To make
these considerations more concrete, we apply them to a mesoscopic
scattering detector similar to that considered in Ref.
\onlinecite{Pilgram}, identifying precise conditions and
symmetries needed to reach the quantum limit.  We find that the
required symmetries are most easily understood if one considers
the scattering detector in terms of information; these symmetries
are not the same as those usually considered in mesoscopic
systems. For example, we find that time reversal symmetry is not
necessary for reaching the quantum limit.  We also find that,
surprisingly, an adiabatic point contact \cite{GlazmanQPC} system
remains a quantum limited detector even for voltages large enough
that several channels contribute to transport and that the energy
dependence of scattering is important; previous studies
\cite{Gurvitz, Stodolsky, KorotkovQPC} have only shown that the
quantum limit is achieved in the small voltage regime.  Our
results for the mesoscopic scattering detector are complementary
to those obtained in Ref. \onlinecite{Pilgram}.

\section{General Conditions}

\subsection{Model and Derivation of the Quantum Limit}

We start by considering a generic system consisting of a qubit
(i.e.~a two-level system described as a spin $1/2$) coupled to an
arbitrary detector.  The system Hamiltonian is $H = H_{\rm qubit}
+ H_{\rm detector} + H_{\rm int}$, where $H_{\rm qubit}  =
-\frac{1}{2} \Omega \sigma_z$, $H_{\rm int}  =  A \sigma_z Q$, and
we leave $H_{\rm detector}$ unspecified.  $Q$ is the detector
``input" operator which couples to the qubit, while $A$
characterizes the strength of the qubit-detector coupling. Mixing
effects, where the detector causes transitions in the qubit, are
neglected by taking $[H_{\rm int},H_{\rm qubit}]=0$; such effects
always cause a deviation from the quantum limit. We work in the
weak-coupling regime ($A \ra 0$), and can thus use linear response
theory to describe the output of detector. Taking $I$ to be the
detector observable that is measured (i.e.~the ``output"
operator), one has to lowest order in $A$:
\begin{equation}
    \langle I(t) \rangle = \langle I(t) \rangle_{\rho_0} +
    A \lambda \langle \hat{\sigma}_z(t) \rangle_{\rho_Q}
    \label{LinResp}
\end{equation}
where the zero-frequency linear-response coefficient (or ``forward
gain") $\lambda$ is given by
\begin{eqnarray}
    \lambda  & \equiv &
        \frac{-i}{\hbar}
        \int_{0}^{\infty} d \tau
                \langle \left[ I(\tau), Q(0) \right]
                \rangle_{\rho_0} \\
        & = &
        \frac{2}{\hbar} \textrm{Im} \int^\infty_{0} d \tau \langle
            I(\tau) Q(0) \rangle_{\rho_0}
            \label{LambdaDefn}
\end{eqnarray}
Here, $\rho_0$ is the initial density matrix of the detector, and
$\rho_Q$ is the initial density matrix of the qubit.  We have
assumed that the qubit splitting frequency $\Omega$ is much
smaller than the rate which characterizes the detector, which
allows us to approximate the detector's response to the qubit as
instantaneous.  Alternatively, one can restrict attention to the
case where the qubit is in a $\sigma_z$ eigenstate, and thus
$\langle \sigma_z(t) \rangle$ is time independent. The operators
on the RHS in the above equation evolve in the Heisenberg picture
generated by $H_0 = H_{\rm qubit} + H_{\rm detector}$.

Next, we connect the detector noise in the output operator $I$ and
input operator $Q$ to, respectively, the measurement rate
$\Gamma_{\rm meas}$ and the dephasing rate $\Gamma_{\varphi}$.
Defining the fluctuating part of an operator $A$ as $\widetilde{A}
= A - \langle A \rangle_{\rho_0}$, the required zero-frequency
noise correlators are given by: \bmlett \label{Corrs}
\begin{eqnarray}
    S_I & = & 2 \int_{-\infty}^{+\infty} dt
        \langle \widetilde{I}(t) \widetilde{I}(0)
        \rangle_{\rho_0} \nonumber \\
        & = &
        4 \pi \hbar \sum_{i,f} P_i \delta(E_i - E_f) |\widetilde{I}_{if}|^2
        \label{SISpect}\\
    S_Q & = & 2 \int_{-\infty}^{+\infty} dt
        \langle  \widetilde{Q}(t) \widetilde{Q}(0)
        \rangle_{\rho_0} \nonumber \\
        & = &
        4 \pi \hbar \sum_{i,f } P_i \delta(E_i - E_f) |\widetilde{Q}_{if}|^2
        \label{SQSpect}\\
    S_{IQ} & = & 2 \int_{-\infty}^{+\infty} dt
        \langle \widetilde{I}(t) \widetilde{Q}(0)
        \rangle_{\rho_0} \nonumber \\
        & = &
        4 \pi \hbar \sum_{i,f} P_i \delta(E_i - E_f) (\widetilde{I}_{if})(\widetilde{Q}_{fi})
    \label{SIQ}
\end{eqnarray}
\emlett %
Here, we use the short hand $O_{if} = \langle i | O | f \rangle$,
where $|i \rangle$,$|f \rangle$ are eigenstates of $H_{\rm
detector}$ with energies $E_i,E_f$.  The probability $P_i$ is
defined as $\langle i | \rho_0 | i \rangle$; we assume that
$\rho_0$ is diagonal in the basis of eigenstates. Taking the
detector noise to be Gaussian, the standard expressions for the
dephasing rate $\Gamma_{\varphi}$ and measurement rate
$\Gamma_{\rm meas}$ are given by: \cite{MakhlinRMP}
\begin{eqnarray}
    \Gamma_{\varphi} & = & \frac{A^2}{\hbar^2} S_Q \label{Rates}
    \hspace{1 cm}
    \Gamma_{\rm meas}
    = \frac{A^2 \lambda^2}{ S_I}
\end{eqnarray}

We briefly review the origin of Eqs. (\ref{Rates}).  The dephasing
rate describes the measurement-induced decay of the off-diagonal
elements of the qubit density matrix. It can be derived by looking
at the decay at long times of the phase correlator $V(t) = \langle
\sigma_+(t) \sigma_{-}(0) \rangle$, where $\sigma_{+}$
($\sigma_{-}$) is the spin raising (lowering) operator:
\begin{eqnarray}
    V(t) & = & \Big\langle \exp \left[ -i \int_{0}^t dt'
    \left( \Omega + 2 A Q(t') / \hbar \right) \right]
    \Big\rangle \label{FirstDephEq} \\
    & \simeq & e^{-i \widetilde{\Omega}  t} \exp \left(
        \frac{-2 A^2}{\hbar^2} \int_0^t dt_1 \int_0^t dt_2 \langle
        \widetilde{Q}(t_1) \widetilde{Q}(t_2) \rangle \right) \nonumber \\
        & \ra & e^{-i \widetilde{\Omega} t}
        e^{-\Gamma_{\varphi} t}
\end{eqnarray}
Here, $\widetilde{\Omega} = \Omega + 2 A \langle Q
\rangle_{\rho_0} / \hbar$.

The measurement rate describes how long the measurement must be on
before the signal associated with the two qubit states can be
distinguished from the noise in $I$.  The quantity of interest is
the time-integral of the detector output, $m(t) = \int_0^t dt'
I(t') $.  One needs that the distributions of $m(t)$ corresponding
to the two different qubit states (i.e.~$p(m(t)|\ua)$ and
$p(m(t)|\da)$) be statistically distinguishable. Assuming Gaussian
distributions, distinguishability is defined as:
\begin{equation}
    \langle m(t)\rangle_{\ua} - \langle m(t) \rangle_{\da} \geq
        \sqrt{2} \left( \sigma_{\ua}(t) + \sigma_{\da}(t)\right),
        \label{MR1}
\end{equation}
where $\sigma$ denotes the variance of the distribution, and the
$\sqrt{2}$ factor is included in order to make the final upper
bound on $\chi$ unity.  Using Eq.~(\ref{LinResp}) for $\langle
I(t) \rangle$, and letting $\tau_{\rm meas} = 1 / \Gamma_{\rm
meas}$, the condition becomes:
\begin{equation}
    2 A \lambda \tau_{\rm meas} \geq
    2\sqrt{2} \cdot \sqrt{\left(\frac{1}{2}S_{II}\right) \tau_{\rm meas}
    },
    \label{MR2}
\end{equation}
which directly yields the expression in Eq.~(\ref{Rates}) for
$\Gamma_{\rm meas}$.  Note that we have taken $\sigma_{\ua} =
\sigma_{\da}$ in the last step; this is sufficient to obtain the
leading order expression for $\Gamma_{\rm meas}$.

To relate $\Gamma_{\varphi}$ and $\Gamma_{\rm meas}$, we first
note that the righthand sides of Eqs.~(\ref{SISpect})-(\ref{SIQ})
implicitly define an inner product (i.e., interpret the matrix
elements $\{\tilde{I}_{if} \}$ and $\{\tilde{Q}_{if} \}$ as
defining vectors). The Schwartz inequality then immediately
yields:
\begin{equation}
    S_I S_Q \geq |S_{IQ}|^2
        = \hbar^2 (\lambda - \lambda')^2 + (\textrm{Re } S_{IQ})^2
\label{SchwartzI}
\end{equation}
where we have introduced the reciprocal response coefficient (or
``backwards gain") $\lambda'$:
\begin{equation}
\lambda' \equiv \frac{2}{\hbar} \textrm{Im} \int^\infty_{0} d \tau
\langle \hat{Q}(\tau) \hat{I}(0) \rangle_{\rho_0}
\end{equation}
$\lambda'$ would describes the response of $\langle Q(t)\rangle$
to a perturbation which couples to the operator $I$.  Note that as
$\lambda$ and $\lambda'$ are defined in terms of commutators, we
may substitute $I \ra \widetilde{I}$, $Q \ra \widetilde{Q}$ in
their definitions.  General stability considerations lead to the
condition $\lambda \lambda' \leq 0$. Using Eqs.~(\ref{Rates}), we
thus have:
\begin{equation}
    \frac{ \Gamma_{\rm meas} }{ \Gamma_{\varphi} } =
        \frac{\hbar^2 \lambda^2}{S_Q S_I} \leq
        \frac{\hbar^2 \lambda^2}
        {\hbar^2 (\lambda - \lambda')^2 + (\textrm{Re} S_{IQ})^2 }
        \leq 1 \label{FinalInEq}
\end{equation}
The best one can do is measure the qubit as quickly as one
dephases it \cite{TwoFactorNote}.  Note that this derivation only
requires the validity of linear response and the weak-coupling
approximations which give rise to Eqs.~(\ref{Rates}); very little
is specified of the detector.  Similar derivations of the quantum
limit are presented in Refs. \onlinecite{Averin} and
\onlinecite{Braginsky}.

The inequality of Eq.~(\ref{FinalInEq}) is in many ways
intuitively reasonable.  Both dephasing and measurement involve
entangling the state of the qubit with states in the detector.  In
principle, there may be degrees of freedom in the detector which
become entangled with the qubit {\it without} providing any
detectable information in a measurement of $\langle I \rangle$;
any such entanglement would lead to $ \Gamma_{\varphi} >
\Gamma_{\rm meas}$.  More precisely, imagine that when the
measurement is initially turned on, the system is in a product
state:
\begin{equation}
    | \psi(t=0) \rangle = \frac{1}{\sqrt{2}} \Big(
        | \ua \rangle + | \da \rangle \Big)
        \otimes | D \rangle,
\end{equation}
where $|D\rangle$ is the initial state of the detector, and $| \ua
\rangle$,$| \da \rangle$ denote qubit $\sigma_z$ eigenstates.  At
some later time $t$, the state of the system may be written as:
\begin{equation}
    | \psi(t) \rangle = \frac{1}{\sqrt{2}} \Big(
        | \ua \rangle \otimes | D_\ua (t) \rangle + | \da \rangle
        \otimes | D_\da (t) \rangle \Big),
    \label{HeuristicPsi}
\end{equation}
To say that we have measured the state of the system implies that
the states $| D_{\ua} (t) \rangle$ and $| D_{\da} (t) \rangle$ are
distinguishable; to say that the qubit has been dephased only
implies that the detector states $| D_{\ua} (t) \rangle$ and $|
D_{\da} (t) \rangle$ are orthogonal.  While distinguishability
implies orthogonality, the opposite is not true; thus, in general,
$\Gamma_{\varphi} > \Gamma_{\rm meas}$. Note that in this
formulation, the dephasing rate will be related to the overlap
between the two detector states:
\begin{equation}
    |\langle D_{\ua} (t) | D_{\da} (t) \rangle| \simeq e^{- \Gamma_{\varphi}
    t}
    \label{PhenDephasing}
\end{equation}

\subsection{Necessary Conditions for Reaching the Quantum Limit}

We have thus seen that on a heuristic level, reaching the quantum
limit requires that the detector have no ``extraneous" degrees of
freedom which couple to the qubit.  Equivalently, all information
on the state of the qubit residing in the detector should be
accessible in a measurement of $\langle I \rangle$.  The virtue of
the derivation presented in the last subsection is that these
statements can be given a precise meaning.  One sees that three
conditions are necessary to reach the quantum limit: (i) the
Schwartz inequality of Eq.~(\ref{SchwartzI}) must be optimized,
(ii) the cross-correlator $\textrm{Re }S_{IQ}$ must vanish, and
(iii) the backwards gain $\lambda'$ must vanish. Conditions (i)
and (ii) can be succinctly re-expressed as a single condition,
leading to the following necessary and sufficient requirements:
\begin{equation}
        \{ \forall i,f | P_i \neq 0, E_f =
        E_i\},
        \langle f | \widetilde{I} | i \rangle   =
        i \CC \langle f | \widetilde{Q} | i \rangle \label{BigCondition}
\end{equation}
\begin{equation}
        \lambda' \equiv \frac{2}{\hbar} \textrm{Im} \int^\infty_{0} d \tau
\langle \hat{Q}(\tau) \hat{I}(0) \rangle_{\rho_0} = 0
    \label{GainCondition}
\end{equation}
Here, $\CC$ is a real number which is independent of the detector
eigenstates $|i\rangle$ and $|f \rangle$.
Eqs.~(\ref{BigCondition}) and (\ref{GainCondition}) are central
results of this paper.  The first of these equations expresses the
fact that to reach the quantum limit, there must be a close
similarity between the detector's input and output operators-- as
far as the zero-frequency noise correlators are concerned, {\it
the operators $I$ and $Q$ must be proportional to one another}.
This required similarity between the detector input and output is
a formal expression of the intuitive idea that a quantum limited
detector has no ``extraneous" internal degrees of freedom.   The
second condition, Eq.~(\ref{GainCondition}), expresses the fact
that a quantum-limited detector must have a strong intrinsic
directionality which discriminates between the input and output.
The output operator is influenced by behaviour at the input, but
not vice-versa.  This requirement is consistent with our tacit
assumption that the quantity $\langle I \rangle$ can be measured
without problems.  To measure $I$, one needs to introduce a
coupling in the Hamiltonian to $I$; the vanishing of $\lambda'$
implies that this additional coupling will not contribute to
$\langle Q(t) \rangle$, and thus cannot further dephase the qubit
(c.f.~Eq.~(\ref{FirstDephEq})).

On a technical level, Eq.~(\ref{BigCondition}) follows from the
optimization of the Schwartz inequality and the requirement that
$\textrm{Re }S_{IQ} = 0$ (i.e.~conditions (i) and (ii) above). The
vanishing of $\lambda'$ (Eq.~(\ref{GainCondition})) can be
interpreted in terms of causality.  To see this, we first
introduce the frequency-dependent cross-correlator $S_{IQ}(E)$:
\begin{eqnarray}
    S_{IQ}(E) & = &
    2 \int_{-\infty}^{\infty} dt
        \langle \widetilde{I}(t) \widetilde{Q}(0) \rangle_{\rho_0} e^{i E t / \hbar}
        \nonumber \\
        & = & 4 \pi \hbar \sum_{i,f\neq i} P_i \delta(E + E_i - E_f)
        \widetilde{I}_{if} \widetilde{Q}_{fi}. \label{SIQFreq}
\end{eqnarray}
We may use this to write:
\begin{eqnarray}
    \lambda (\lambda')  = &&  \frac{1}{2 \hbar}  \Bigg(
        + (-)  \textrm{Im }\left[ S_{IQ}(0) \right] \nonumber \\
         && -
        \frac{1}{\pi} {
         P} \int_{-\infty}^{\infty} dE
        \frac{ \textrm{Re } \left[ S_{IQ}(E) \right] }
          { E } \Bigg) \label{LambdaDecomp}
\end{eqnarray}
If $\lambda' = 0$, it follows from the above that at $E=0$, the
imaginary part of $S_{IQ}(E)$ coincides with the Hilbert transform
of the real part of $S_{IQ}(E)$:
\begin{equation}
        \textrm{Im }\left[ S_{IQ}(E) \right] \Bigg|_{E=0} =
        \Bigg(-
        \frac{1}{\pi} {
         P} \int_{-\infty}^{\infty} dE'
        \frac{ \textrm{Re } \left[ S_{IQ}(E') \right] }
          { E' - E } \Bigg) \Bigg|_{E=0}
        \label{WeakCausal}
\end{equation}
If this held for all $E$, it would follow from the Titchmarsh
theorem \cite{TitRef} that $S_{IQ}(t) = \langle \widetilde{I}(t)
\widetilde{Q}(0) \rangle_{\rho_0} $ is causal: it would vanish for
$t<0$. This would clearly be sufficient to satisfy
Eq.~(\ref{GainCondition}). More generally, the vanishing of
$\lambda'$ only requires the weaker condition of
Eq.~(\ref{WeakCausal}).

\subsection{The Quantum Limit and Information Theory}

We close this section by formalizing the connection between the
quantum limit and information.  A deviation from the quantum limit
(i.e.~$\chi < 1$) implies the existence in the detector of
``missing information" regarding the state of the qubit,
information which is not revealed in a measurement of $\langle I
\rangle$.  The dephasing rate thus corresponds to what the
measurement rate would be {\it if} we could make use of all the
available information. This notion can be quantified by borrowing
a concept from quantum information theory, the accessible
information \cite{Levitin,Fuchs,Nielsen,Vedral}. To define this,
note first that if we choose a specific detector quantity (or set
of  quantities) $Y$ to measure (described by, e.g., a set of
commuting observables), we can think of our system as a noisy
classical communication channel.  The two possible inputs to the
channel are the qubit states $| \ua \rangle$ and $| \da \rangle$;
interaction with the detector for a time $t$ then leads to two
corresponding detector states $| D_{\ua} (t) \rangle$ and $|
D_{\da}(t) \rangle$ (c.f.
Eq.~(\ref{HeuristicPsi})).\cite{PureCaveat} Finally, the outputs
from the channel are the outcomes of the measurement of $Y$. The
``noise" here is a result of the intrinsic uncertainties of $Y$ in
the states $| D_{\ua}(t) \rangle$ and $| D_{\da}(t) \rangle$; the
output will thus be described by the conditional probability
distributions $p(y|\ua),p(y|\da)$ determined by these states,
where $y$ represents possible outcomes of the measurement. Letting
$\bar{p}(y) = [p(y|\ua) + p(y|\da)]/2$, the mutual information $R$
of this channel is \cite{InfoRef}:
\begin{equation}
    R[Y] = H[ \bar{p}(y) ] - \frac{1}{2} \Big(
        H[ p(y| \ua) ] + H[ p(y | \da) ] \Big) \label{REqn}
\end{equation}
where $H[p(y)]$ is the Shannon information entropy associated with
the distribution $p$:
\begin{equation}
    H[p(y)] = -\sum_{y_i} p(y_i) \log(p(y_i))
\end{equation}
Note that we have chosen to equally weight our two inputs to the
channel.  Assuming that this choice is optimal, Shannon's noisy
channel coding theorem implies that $R[Y]$ is the maximum rate at
which messages can be reliably transmitted down the channel by
modulating the state of the qubit and making measurements of $Y$.
\cite{InfoRef} Alternatively, $R[Y]$ may be considered as being
related to a generalized measurement rate describing the chosen
measurement $Y$.  For example, if the distributions $p(y(t) |
\ua)$ and $p(y(t) | \da)$ are Gaussian, one finds that at small
times (i.e.~before the two distributions are well separated):
\begin{equation}
    R[Y]_{\rm Gaussian} = \frac{1}{8}
    \frac{
        \left(
            \langle y(t) \rangle_{\ua} - \langle y(t)
            \rangle_{\da}
        \right)^2
        }
        {   \sigma_{\ua}(t)\sigma_{\da}(t)  }
\end{equation}
This corresponds to our definition of the measurement rate,
c.f.~Eqs.~(\ref{MR1}) and (\ref{MR2}).  We thus have a new way to
interpret the measurement rate $\Gamma_{\rm meas}$:  given that
one is monitoring $\langle I \rangle$, $\Gamma_{\rm meas}$
represents the maximum rate at which information can be sent to
the detector by modulating the qubit.

The quantum mechanical accessible information $\mathcal{I}$ is now
defined by maximizing the mutual information $R[Y]$ over all
possible measurement schemes $Y$. Remarkably, for the case
considered here (where the detector is described by a pure state)
it can be calculated exactly \cite{Levitin}; a simplified proof is
presented in Appendix A, where we also demonstrate that there are
several possible optimal measurement schemes.  Letting $|\langle
D_{\ua} (t) | D_{\da} (t) \rangle|^2 = \cos^2(\alpha(t))$, we
have:
\begin{eqnarray}
    {\mathcal I} = \max_{\{Y\}} R   = && \frac{1}{2} \Big[
        (1 + \sin \alpha(t) ) \log (1 + \sin \alpha(t) ) +
        \nonumber \\
        && (1 - \sin \alpha(t) ) \log (1 - \sin \alpha(t) ) \Big]
    \label{Accessible}
\end{eqnarray}
This expression corresponds to having equally weighted our two
input states, as we did in Eq.~(\ref{REqn}); one can check that
this choice maximizes $\mathcal I$.  At small times
($\Gamma_{\varphi} t \ll 1$), comparison against
Eq.~(\ref{PhenDephasing}) yields $\alpha(t) \ra 0$, and we have:
\begin{equation}
     {\mathcal I} \simeq \alpha(t)^2 = \Gamma_{\varphi} t
    \label{WeakIA}
\end{equation}
As expected, the growth of the accessible information is
determined by the dephasing rate.  Achieving $\chi=1$ thus implies
that the rate that we actually obtain information, $\Gamma_{\rm
meas}$, coincides with the growth of the total accessible
information. Thus, there is no ``missing" information in the
detector.  We can also think of Eqs. (\ref{Accessible}) and
(\ref{WeakIA}) as providing an alternate route for deriving the
quantum limit inequality $\Gamma_{\varphi} \geq \Gamma_{\rm
meas}$, i.e.:
\begin{equation}
    R[Y] \simeq \Gamma_{\rm meas} t \leq {\mathcal I} \simeq
    \Gamma_{\varphi} t
\end{equation}
The utility of thinking about back action effects and the quantum
limit in terms of information will become clear in the next
section, where we discuss the mesoscopic scattering detector. Note
also that the relation between information and state disturbance
has been studied in a slightly different context by Fuchs et al.
\cite{Fuchs}

\section{Mesoscopic Scattering Detector}

To make the preceding discussion more concrete, we now consider
the status of the quantum limit in a slightly less general
detector set-up, the mesoscopic scattering detector considered in
Ref. \onlinecite{Pilgram}.  We determine the conditions needed to
reach the quantum limit of detection by directly applying the
general conditions derived in the last section, namely the
proportionality condition of Eq.~(\ref{BigCondition}), and the
causality condition of Eq.~(\ref{GainCondition}).  This is in
contrast to Ref. \onlinecite{Pilgram}, which developed conditions
needed for the quantum limit by directly calculating
$\Gamma_{\varphi}$ and $\Gamma_{\rm meas}$.  We explicitly show
that a violation of Eq.~(\ref{BigCondition}) implies the existence
of unused information in the detector, information which is not
extracted in the measurement process.

The detector here is a two terminal scattering region (see Fig. 1)
characterized by a scattering matrix $s$.  Taking the contact to
both the right and left reservoirs to have $N$ propagating
transverse modes, $s$ will have dimension $2 N$. The output
operator of the detector $I$ is simply the current through the
region; the state of the qubit alters $\langle I \rangle$ by
modulating the potential in the scattering region.  Note that
while we focus on the limit of a weak coupling between the qubit
and detector, so that the linear response approach of the previous
section is valid, we do not assume that the voltage is small
enough that $\langle I \rangle \propto V$. \cite{NonlinearNote}
The mesoscopic scattering detector describes the setup used in two
recent ``which path" experiments \cite{Buks,Heiblum}. These
experiments used a quantum point contact to detect the presence of
an extra electron in a nearby quantum dot.  As the dot was
imbedded in an Aharanov-Bohm ring, the dephasing induced by the
measurement could be studied directly.

\begin{figure}
\includegraphics[width=6 cm]{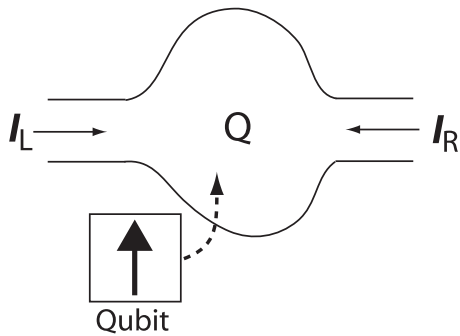}
\caption{\label{DetectorPic} Schematic of the mesoscopic
scattering detector, in which the current through a phase coherent
scattering region is used to detect the qubit.  $Q$ denotes the
charge in the scattering region, while $I_R$ ($I_L$) is the
current in the right (left) contact.}
\end{figure}

We start by considering the simplest situation, also considered in
Ref. \onlinecite{Pilgram}, where the state of the qubit provides a
uniform potential change in the scattering region. In this case
the input operator $Q$ is the {\it total} charge in the scattering
region. Unlike Ref. \onlinecite{Pilgram}, we do not explicitly
consider the effects of screening here.  Within an RPA scheme,
consideration of such effects allows an explicit calculation of
the qubit-detector coupling strength $A$, but does not result in
any other changes over a non-interacting approach. In the weak
coupling regime, the particular value of $A$ does not affect the
approach to the quantum limit.

Letting $a_{\alpha n}^{\dagger}(E)$ represent the creation
operator for an incident wave in contact $\alpha=L,R$, transverse
mode $n$, and at energy $E$, the detector current operator for
contact $\alpha$ takes the form \cite{ButtikerReview}:
\begin{eqnarray}
    I_{\alpha} & = &  \frac{e}{h} \int dE \int dE'  \sum_{\beta, \gamma = L,R}
    \sum_{n,m=1}^N \nonumber \\
        &&
    \left[ a_{\beta n}^{\dagger}(E)
        A_{\beta n, \gamma m}(\alpha; E,E')
        a_{\gamma m}^{\pd}(E') \right]
    \label{IDefn}
\end{eqnarray}
\begin{equation}
    A_{\beta n, \gamma m}(\alpha; E,E')  =
    \delta_{\beta \gamma} \delta_{\alpha \beta} \delta_{n m} -
        \left(
        \left[ s_{\alpha \beta}(E) \right]^{\dagger}
        s_{\alpha \gamma}(E')
        \right)_{nm}
\end{equation}
A positive current corresponds to a current incident on the
scattering region; note that throughout this section, we neglect
electron spin for simplicity. The total charge $Q$ in the
scattering region may be defined in terms of the total current
incident on the scattering region-- in the Heisenberg picture,
$\partial_t Q(t) = I_L(t) + I_R(t)$ . One obtains:
\begin{eqnarray}
    Q
     & = &   e \int dE \int dE'  \sum_{\beta, \gamma =
     L,R} \nonumber \\
    && \left[ a_{\beta n}^{\dagger}(E) \NN_{\beta n, \gamma m}^{\pd}(E,E')
        a_{\gamma m}^{\pd}(E') \right]
    \label{QDefn}
\end{eqnarray}
\begin{eqnarray}
    \NN(E,E+ \hbar \omega) & = & \frac{1}{2 \pi i}
    \left[ s^{\dag}(E) \frac{s(E+ \hbar \omega)-s(E)}{\hbar \omega}
    \right].
    \label{NNDefn}
\end{eqnarray}
In the limit where $\omega \ra 0$, $\NN(E,E+\hbar \omega)$ reduces
to the well-known Wigner-Smith delay time matrix:
\begin{equation}
    \NN(E) = \frac{1}{2 \pi i}
    \left[ s^{\dag}(E) \frac{d}{dE} s(E) \right]
\end{equation}
Finally, the assumption that the qubit couples to the total charge
in the scattering region is equivalent to assuming that the
potential it creates is smooth in the WKB sense.  We can use the
fact that the sensitivity of the scattering matrix $s$ to a global
change of potential in the scattering region is the same as its
sensitivity to energy.  Thus, the linear response coefficient
$\lambda$ has the form:
\begin{eqnarray}
    \lambda & = & - \frac{e^2}{h} \int_{\mu_R}^{\mu_L} d \ve
          \frac{d}{d \ve}
          \left[
            \textrm{tr }
             s_{LR}^{\dagger}(\ve)
             s_{LR}(\ve) \right] \nonumber \\
        & = &  - \frac{e^2}{h} \int_{\mu_R}^{\mu_L} d \ve \sum_j
          \frac{d T_j (\ve)}{d \ve} ,
        \label{SCLambda}
\end{eqnarray}
where the $T_j$ are the transmission eigenvalues of the system.
Without loss of generality, we have assumed that our detector is
biased such that the chemical potential of the left reservoir is
greater than that of the right reservoir: $\mu_L - \mu_R = e|V|$;
we also consider the limit of zero temperature.

\subsection{Single Channel Case}

Given these definitions, we can now turn to
Eqs.~(\ref{BigCondition}) and (\ref{GainCondition}) and ask what
is required of the scattering matrix $s$ in order to reach the
quantum limit. We first focus on the case $N=1$, where there is a
single propagating mode in both contacts.  The scattering matrix
$s$ is thus $2 \times 2$, and may be written as:
\begin{equation}
    s(E) =
    \left(
        \begin{array}{cc}
          s_{LL} & s_{LR} \\
          s_{RL} & s_{RR} \\
        \end{array}
    \right)
    =
    \left(
        \begin{array}{cc}
          \sqrt{R} e^{i \beta} & \sqrt{T} e^{i \varphi'} \\
          \sqrt{T} e^{i \varphi} & -\sqrt{R} e^{ i (\varphi + \varphi' - \beta)} \\
        \end{array}
    \right)
    \label{SDecomp}
\end{equation}
where $R = 1-T$.  At zero temperature, the detector is described
by a single many-body state $|i \rangle$ in which all incident
states in lead $\alpha$ with $E < \mu_{\alpha}$ are occupied, and
all other incident states are unoccupied:
\begin{equation}
    | i \rangle =
    \left( \Pi_{E_L \leq \mu_L} a^{\dag}_L(E_L) \right)
        \left( \Pi_{E_R \leq \mu_R} a^{\dag}_R(E_R) \right)
         | \textrm{vac} \rangle
\label{StateDefn}
\end{equation}

First, we consider the causality condition of
Eq.~(\ref{GainCondition}) which requires that the backwards gain
$\lambda'$ vanishes.  As we know the initial state of the detector
and have explicit expressions for $I$ and $Q$, we can directly
evaluate the function $S_{IQ}(E)$ appearing in Eq.~(\ref{SIQFreq})
in terms of $s$.  A direct calculation can be performed to show
that:
\begin{eqnarray}
        \int_{-\infty}^{\infty} dE
        \frac{ \textrm{Re } \left[ S_{IQ}(E) \right] }
          { E }         & = &
        \int_{-\infty}^{\infty} dE
        \frac{ \textrm{Re } \left[ F(E) \right] }
          { E } \label{CausalOne} \\
          \textrm{Im } \left[ S_{IQ}(0) \right] & = &
          \textrm{Im } \left[ F(0) \right] \label{CausalTwo}
          \end{eqnarray}
where, letting $t \equiv s_{RL}$,the function $F(E)$ is defined
as:
\begin{equation}
    \label{Gfunc}
    F(E) = -i \frac{e^2 }{2 \pi} \int_{\mu_R}^{\mu_L} dE' t^*(E') \left(
        \frac{t(E'+E) - t(E')}{E} \right)
\end{equation}
Note that Eqs.~(\ref{CausalOne}) and (\ref{CausalTwo}) are
independent of whether $I$ is take to be $I_L$, $I_R$, or a linear
combination of the two.  Now, causality dictates that the
scattering matrix $s$ is analytic in the upper half complex plane,
and thus so is the function $F(E)$.  The real and imaginary parts
of $F$ are thus related by a Hilbert transform, and
Eqs.~(\ref{WeakCausal}), (\ref{CausalOne}) and (\ref{CausalTwo})
imply that $\lambda'=0$ for the scattering detector {\it
irrespective of the choice of s}. Thus, the causality properties
of the scattering matrix $s$ ensure that one of the conditions
necessary for reaching the quantum limit is always satisfied. Note
that substituting these expressions for $S_{IQ}(E)$ in
Eq.~(\ref{LambdaDecomp}) does indeed yield the expected form of
$\lambda$ (Eq.~(\ref{SCLambda})).  It is also useful to note that
gauge invariance can be used to directly establish\cite{Levitov}
$\lambda'=0$ . The essence of the argument is that a coupling to
the current (i.e.~$H_{\rm int} = A \sigma_z I(x=0)$) is equivalent
to introducing a local vector potential. The gauge transformation
which removes this term will only modify the transmission phases
in the scattering matrix $s$ (i.e.~$\phi$ and $\phi'$) in an
energy-independent manner. Using Eq.~(\ref{QDefn}), one can check
that $\langle Q \rangle$ is independent of energy-independent
phase changes; thus $\lambda'=0$.

Next, we turn to the condition given in Eq.~(\ref{BigCondition}),
which requires a certain proportionality between $\widetilde{I}$
and $\widetilde{Q}$ in order to reach the quantum limit.  Given
the state $|i\rangle$ which describes the detector
(Eq.~(\ref{StateDefn})), the only matrix elements of $I$ and $Q$
which contribute to the zero frequency noise correlators (c.f.~Eqs
(\ref{Corrs})) involve energy-conserving transitions where a
scattering state incident from the left reservoir is destroyed
while a scattering state incident from the right reservoir is
created.  Since these transitions require an occupied initial
state and an unoccupied final state, they can only occur in the
energy interval $\mu_R < E < \mu_L$.  We are thus interested in
the coefficients of the operators $a_{R}^{\dagger}(E)
a_{L}^{\pd}(E)$ appearing in the expansion of $I$ and $Q$ in this
energy interval.  The proportionality requirement of
Eq.~(\ref{BigCondition}) thus results in a necessary condition on
$s(\ve)$:
\begin{eqnarray}
    \forall E \epsilon [\mu_R,\mu_L],
    \hspace{0.2 cm}
    \left[s_{LR}
\right]^{*}(E) s_{LL}^{\phantom{*}} (E)  =
        i \CC \NN_{RL}(E)
        \label{SCondition}
\end{eqnarray}
where $\CC$ is a real, energy-independent constant.  Using
Eq.~(\ref{SDecomp}), the imaginary and real parts of the above
condition become:
\begin{eqnarray}
    \forall E \epsilon [\mu_R,\mu_L],
    \hspace{0.2 cm}
    \frac{d}{dE} \left(\beta(E) - \phi(E) \right)
        & = &  0 \label{PhaseCondition}
    \\  \frac{  \frac{dT}{dE}(E)  }
        {T(E) (1-T(E))}
       & = &  -\frac{4 \pi}{\CC}  \label{TCondition}
\end{eqnarray}
Similar conditions for reaching the quantum limit for this version
of the scattering detector were first developed in Ref.
\onlinecite{Pilgram} by directly calculating $\Gamma_{\rm meas}$
and $\Gamma_{\varphi}$ (note there is a sign error in Eq.~(7) of
Ref. \onlinecite{Pilgram} which must be corrected to obtain our
Eq.~(\ref{PhaseCondition})). \cite{ConstantNote}  The fulfilling
of these conditions does not correspond to symmetries usually
considered in mesoscopic systems; for example, as we will show,
the presence of time-reversal symmetry is not a necessary
requirement.  Instead, the conditions of Eqs.
(\ref{PhaseCondition}) and (\ref{TCondition}) correspond directly
to the requirement that there be no missing information in the
detector, information which is not revealed in a measurement of
$\langle I \rangle$.  We demonstrate this explicitly in what
follows.

\subsubsection{Phase Condition}

The first condition (Eq.~(\ref{PhaseCondition})) for reaching the
quantum limit requires that the difference between transmission
and reflection phases in the scattering matrix be constant in the
energy interval defined by the voltage.  If it holds, changing the
state of the qubit will not modulate this phase difference.
Eq.~(\ref{PhaseCondition}) thus constrains information-- it
ensures that the detector does not extract additional information
about the qubit which resides in the relative phase between
transmission and reflection.  Such information is clearly not
revealed in a measurement of $\langle I \rangle$, and would
necessarily lead to additional dephasing over and above the
measurement rate.  In principle, this additional information could
be extracted by performing an interference experiment. To be more
specific, note that the cross-correlator $S_{IQ}$ (c.f.~Eq.
\ref{SIQ}) is given by:
\begin{equation}
    S_{IQ} = i \hbar \lambda + \frac{e^2}{\pi} \int_{\mu_R}^{\mu_L} dE' \left(
        T(1-T) \frac{d}{dE}(\beta - \varphi) \right)
\end{equation}
By definition, the imaginary part of this correlator determines
the linear response coefficient $\lambda$
(c.f.~Eq.~(\ref{LambdaDefn})) associated with measuring $\langle I
\rangle$. In contrast, the {\it real} part of this correlator may
be interpreted as the linear response coefficient associated with
a measurement where one interferes reflected and transmitted
electrons;  the factor of $T(1-T)$ corresponds to the fact that
the magnitude of this signal will be proportional to the amplitude
of both the reflected and transmitted beams.  More explicitly,
consider the Hermitian operator $I_{\rm mod}$ defined by:
\begin{equation}
    I_{\rm mod} = \frac{e}{\hbar} \int_{\mu_R}^{\mu_L} dE
    \left[ i a_{R}^{\dagger}(E) A_{R L}(L; E,E)
        a_{L}^{\pd}(E)  + h.c.  \right]
    \label{IModDefn}
\end{equation}
If one were to now measure $I_{\rm mod}$, the corresponding linear
response coefficient $\lambda_{\rm mod}$ is precisely the real
part of $S_{IQ}$ (this can be seen by comparing
Eqs.~(\ref{IModDefn}) and (\ref{IDefn}) ). The fact that
additional information on the state of the qubit is available in
the expectation $\langle I_{\rm mod} \rangle$ implies that the
qubit is entangling with the detector faster than the measurement
rate associated with $\langle I \rangle$.  This remains true even
if one does not explicitly extract this information, as was
demonstrated recently in the experiment of Sprinzak et. al.
\cite{Heiblum}

Stepping back, we see that the general condition $\textrm{Re }
S_{IQ} = 0$ (i.e.~the required factor of $i$ on the RHS of Eq.
\ref{BigCondition}) needed to reach the quantum limit directly
corresponds to the requirement of no ``missing" information
discussed in the previous subsection.  In general, a non-vanishing
$\textrm{Re } S_{IQ}$ implies that additional information about
the qubit's state could be obtained by simultaneously measuring
another quantity in addition to $I$ (e.g., in our case, the
quantity $I_{\rm mod}$).

Note that in the scattering detector, the symmetry required to
ensure that Eq.~(\ref{PhaseCondition}) holds (i.e.~that the phases
$\beta$ and $\phi$ coincide) is not one that is usually considered
in mesoscopic systems. In particular, {\it the presence of
time-reversal symmetry is not {\it necessary} to fulfilling the
condition of Eq.~(\ref{PhaseCondition})}; time-reversal symmetry
only implies that $\varphi = \varphi'$, and specifies nothing on
the relation between $\varphi$ and $\beta$. However, as pointed
out in Ref. \onlinecite{Pilgram}, a {\it sufficient} condition for
achieving Eq.~(\ref{PhaseCondition}) is that one has parity
symmetry, that is {\it both} time-reversal symmetry and left-right
inversion symmetry (the latter condition implies that the two
reflection phases in $s$ are identical).\cite{SymmNote} Note that
this is not a necessary condition.  We see that the required
symmetry here is best understood as being related to information.

\subsubsection{Transmission Condition}

We now turn to the second condition (Eq.~(\ref{TCondition}))
needed to have the scattering detector reach the quantum limit, a
condition which constrains the energy dependence of the
transmission probability $T$. This condition arises from the
requirement that the proportionality between $I$ and $Q$ needed
for the quantum limit must hold over the entire energy interval
defined by the voltage. In general, energy averaging causes a
departure from the quantum limit-- over sufficiently large
intervals, the operators $I$ and $Q$ look less and less like one
another.  Like Eq.~(\ref{PhaseCondition}), Eq.~(\ref{TCondition})
can also be interpreted as a requirement of no ``missing"
information. Here, the requirement is that energy averaging does
not result in the loss of information about the qubit which is
encoded in the energy dependence of $T$.  While such information
is not obtained in a measurement of $\langle I \rangle$ (which
involves energy averaging, c.f.~Eq.~(\ref{SCLambda})), it could be
obtained if one measured the entire function $\langle I(V)
\rangle$ for $0 \leq |V| \leq \mu_L - \mu_R$.  As discussed, the
presence of any missing information necessarily implies a
departure from the quantum limit.

Interestingly enough, Eq.~(\ref{TCondition}) may be understood
completely classically, even though it formally results from
requiring the proportionality of two quantum operators.  To do so,
we calculate the classical information capacity $R$
(c.f.~Eq.~(\ref{REqn})) corresponding to two different possible
measurements. First, imagine we measure the integrated current $m
= \int_0^t dt' I(t')$, and assume the probability distributions
$p(m| \ua)$ and $p(m| \da)$ are Gaussian. For weak coupling, one
finds for the capacity:
\begin{eqnarray}
    R_{\rm avg} = \Gamma_{\rm meas} t & = & \frac{t}{2 h}
        \frac{
            \left(
                e A \int_{\mu_R}^{\mu_L} d\ve \frac{d T(\ve)}{
                d\ve}
            \right)^2
        }
        {\int_{\mu_R}^{\mu_L} d\ve T(\ve) (1 - T(\ve)) } \\
        & \simeq &
    \frac{(\delta \ve) t}{2 h}
        \frac{
            \left(
                e A \sum_j  \frac{d T(\ve_j)}{d \ve}
            \right)^2
        }
        {\sum_j T(\ve_j) (1 - T(\ve_j)) }
\end{eqnarray}
In the last line, we have discretized the energy integrals
i.e.~partitioned the interval $[\mu_R,\mu_L]$ into equal segments
of length $\delta \ve$.  If we now imagine we could measure each
$m_j = \int_0^t I_j(t)$, where $I_j(t)$ is the contribution to the
current from the $j$th energy interval, a similar calculation
reveals:
\begin{eqnarray}
    R_{\rm tot}
        & = &
    \frac{(\delta \ve) t}{2 h} \sum_j
        \frac{
            \left(
                e A   \frac{d T(\ve_j)}{d \ve}
            \right)^2
        }
        {T(\ve_j) (1 - T(\ve_j)) }
\end{eqnarray}
One can easily check that $R_{\rm tot} \geq R_{\rm avg}$; this
corresponds to the additional information that is generally
available in the energy dependence of $T$.  A necessary and
sufficient condition for ensuring $R_{\rm tot} = R_{\rm avg}$ is
precisely the condition of Eq.~(\ref{TCondition}).  On a purely
classical level, this condition ensures that no information is
lost when one averages over energy.

How can the problems generally posed by energy averaging be
avoided? One possible solution would be to use voltages small
enough that the scattering matrix $s$ can be approximated as being
linear in energy, that is $ e V (dT/dE) \ll 1 $ (this is the
approach of Ref. \onlinecite{Pilgram}). However, as the linear
response coefficient $\lambda$ is given by the energy derivative
of the transmission (c.f.~Eq.~(\ref{SCLambda})), such a small
voltage would imply both a small signal and essentially no gain.
The change in current induced by the qubit, $\Delta I  = \pm A
\lambda $, would be much smaller than the current associated with
the coupling voltage $A$:
\begin{eqnarray}
    \lambda &\simeq & \frac{e^2}{h}
        \left( \frac{dT}{dE} e|V| \right) \ll
    \frac{e^2}{h} \\
    \Gamma_{\rm meas} & \propto & \left(\frac{dT}{dE} eV\right)^2
        \left(\frac{A}{eV}\right) \frac{A}{h}
        \ll \frac{A}{h}
\end{eqnarray}
Even though this smallness of $\lambda$ does not theoretically
affect the approach to the quantum limit, it does severely limit
the detector's practical value-- for very slow measurement rates,
environmental effects on the qubit will become dominant over
backaction effects.

If we now consider finite voltages and fully energy-dependent
scattering, Eq.~(\ref{TCondition}) tells us the condition under
which energy averaging the transmission does not impede reaching
the quantum limit. The solution to Eq.~(\ref{TCondition}) has the
form:
\begin{equation}
    T(E) = \frac{1}{1 + e^{4 \pi (E-E_0) / \CC}}
\end{equation}
This form for $T(E)$ implies that there is no extra information in
the energy dependence of $T$ which is lost upon energy averaging.
Amusingly, Eq.~(\ref{TCondition}) corresponds {\it exactly} to the
energy-dependent transmission of one channel of an adiabatic
quantum point contact \cite{GlazmanQPC}.  The constant $E_0$
represents the threshold energy of the channel (i.e.~the
transverse mode), and the constant $\CC$ is given by:
\begin{equation}
    \CC =  - \frac{ 2 \sqrt{2} \hbar v_F}{\sqrt{d R}}
\end{equation}
where $d$ is the transverse width of the constriction at its
center, and $R$ is the radius of curvature of the transverse
confining potential at the constriction center.

\subsection{Multichannel Case}

We now consider the situation where there are $N$ channels in each
of the two contacts leading to the reservoirs.  It is useful to
write $s$ in terms of its $N$ transmission eigenvalues $T_j(E)$
using the standard polar decomposition:\cite{BeenakkerRMT}
\begin{equation}
    s(E) = \left( \begin{array}{cc}
        s_{LL} & s_{LR} \\
        s_{RL} & s_{RR}
    \end{array} \right) =
    \left( \begin{array}{c}
      U \\
      V \
    \end{array}\right)
    \left(\begin{array}{cc}
      \sqrt{R} & \sqrt{T} \\
      \sqrt{T} & -\sqrt{R} \
    \end{array}\right)
    \left(\begin{array}{c}
      U' \\
      V' \
    \end{array}\right) \label{Polar}
\end{equation}
Here, $U,U',V,V'$ are $N \times N$ energy-dependent unitary
matrices, and $\sqrt{R}$ and $\sqrt{T}$ are diagonal matrices
having entries $\sqrt{1 - T_j(E)}$ and $\sqrt{T_j(E)}$,
respectively.

In the multichannel case, the backwards gain $\lambda'$ again
vanishes irrespective of the details of $s$ as a result of the
analytic properties of $s$. The relevant question then to ask is
what conditions must be satisfied by $s(E)$ so that the
proportionality between $I$ and $Q$ required to reach the quantum
limit (i.e.~Eq.~(\ref{BigCondition}) ) is achieved.  As in the
single-channel case, the relevant matrix elements of $I$ and $Q$
involve destroying a scattering state incident from the left and
creating an equal-energy state describing an incident wave from
the right; the additional complication now is that these
transitions could result in a change of transverse mode.  One thus
needs to examine the coefficients of the operator products $a_{R n
}^{\dagger}(E) a_{L m}^{\pd}(E)$ appearing in the expansion of $I$
and $Q$, in the energy interval $[\mu_R, \mu_L]$.  The
proportionality condition of Eq.~(\ref{BigCondition}) again yields
the requirement that Eq.~(\ref{SCondition}) hold for all energies
in this interval; now, however, both the right and left-hand side
of this equation are $N \times N$ matrices:
\begin{eqnarray}
    \forall E \epsilon [\mu_R,\mu_L],
    \hspace{0.2 cm}
    \left[s_{LR}(E)
\right]^{\dagger} s_{LL}^{\pd} (E)  =
        i \CC \NN_{RL}(E)
        \label{MCSCondition}
\end{eqnarray}
Here, $\CC$ is again an energy-independent real number.  Using the
polar decomposition, one can derive from Eq.~(\ref{MCSCondition})
two necessary matrix conditions which must hold for all energies
in the interval defined by the voltage:
\begin{equation}
    \sqrt{T(E)} \phi_U (E) \sqrt{R(E)} -
    \sqrt{R(E)} \phi_V (E) \sqrt{T(E)}  =  0
    \label{MCPhaseCondition}
\end{equation}
\begin{equation}
    \frac{ \frac{d T}{dE}(E)}
                {T(E) (1 - T(E))}  =  -\frac{4 \pi}{\CC} \times \hat{1}
                \label{MCTCondition}
\end{equation}
These conditions are the multi-channel analogs of
Eqs.~(\ref{PhaseCondition}) and (\ref{TCondition}).  $\hat{1}$
denotes the $N \times N$ unit matrix, and we have introduced the
generalized ``phase-derivative" Hermitian matrices $\phi_U$ and
$\phi_V$:
\begin{eqnarray}
    \phi_U(\ve) & = & -i U^{\dag}(\ve)
        \left[ \frac{d}{dE} U(\ve)\right] \\
    \phi_V(\ve) & = & -i V^{\dag}(\ve) \left[
        \frac{d}{dE} V(\ve) \right]
\end{eqnarray}
These matrices play the role of the energy-derivatives of the
phases $\beta$ and $\phi$ in the single channel case.  Note the
evident asymmetry in Eq.~(\ref{MCPhaseCondition}): the polar
decomposition matrices $U$ and $V$ enter, but the matrices $U'$
and $V'$ do not.  We comment on this in what follows.

\subsubsection{Phase and Channel Mixing Conditions}

The first requirement (Eq.~(\ref{MCPhaseCondition})) places a
stringent requirement on the scattering matrix $s$. Like the
corresponding requirement for the single-channel system, it
ensures that there is no additional information on the state of
the qubit available in measurable changes of scattering phases.
Again,  {\it time-reversal symmetry is not necessary} to have this
condition hold, as time-reversal symmetry only ensures $U=U'$ and
$V = V'$. However, unlike the single-channel case, even the
presence of parity symmetry (i.e.~the combination of both
time-reversal symmetry and left-right inversion symmetry) is {\it
not sufficient} to guarantee that Eq.~(\ref{MCPhaseCondition}) is
satisfied.  The presence of parity symmetry would indeed ensure
$\phi_U = \phi_V$, but as in general $\left[\sqrt{T},\phi_U
\right], \left[\sqrt{R},\phi_U \right] \neq 0$, this is not
enough. In addition to having $\phi_U=\phi_V$, one also generally
needs either that $\phi_U$ is diagonal, meaning that the mode
index (i.e.~transverse momentum) is conserved during scattering,
or that all the transmission eigenvalues $T_j$ are identical.  We
thus see that if the transmissions fluctuate, mode-mixing
(e.g.~the non-conservation of transverse energy) also prevents one
from reaching the quantum limit of detection. This can be
understood from the point of view of information. If the $\phi_U,
\phi_V $ matrices are not purely diagonal, information about the
qubit could be gained by looking at changes in how electrons
incident in a given mode are partitioned into outgoing modes. Such
changes would not be detectable if all channels had the same
transmission.  Note that the matrices $U'$ and $V'$ appearing in
the polar decomposition of $s$ (Eq.~(\ref{Polar})) are irrelevant
to reaching the quantum limit. As each transverse mode is equally
populated with incoming waves in the state $|i \rangle$, there is
no information associated with the preferred mode structure for
incoming waves (i.e. the eigenvectors of $U'$ and $V'$).

\subsubsection{Transmission Condition}

Consider now the condition imposed by Eq.~(\ref{MCTCondition}),
which constrains the form of the transmissions $T_j(\ve)$ of the
detector. Similar to the corresponding condition for the
single-channel system, this requirement ensures that there is no
additional information available in either the energy {\it or}
channel structure of the $\{T_j(\ve)\}$ which is lost upon
averaging. One obtains a necessary form for the transmissions,
similar to what was found in Ref. \onlinecite{Pilgram}:
\begin{equation}
    T_j(E) = \frac{1}{1 + e^{4 \pi (E-E_j) / \CC}}
    \label{TEDep}
\end{equation}
Note that different modes differ from one another only by their
threshold energy $E_j$; the constant $\CC$ is the same for each
mode.  Again, this form for the transmissions \{$T_j(\ve)$\}
corresponds exactly to those expected for a multi-channel
adiabatic point contact.\cite{GlazmanQPC}  The assumption of
adiabaticity implies that transverse energy is conserved.  Thus,
if parity symmetry also holds, we reach the surprising conclusion
that {\it a multi-channel adiabatic point contact remains a
quantum limited detector even if the voltage is large enough that
several modes contribute to transport}. Previous studies have
established that point contact detectors reach the quantum limit
in the limit of small voltages, where the energy-dependence of
scattering can be neglected \cite{Gurvitz,Stodolsky,KorotkovQPC}.
We have shown here that in the adiabatic case, the quantum limit
continues to hold even at voltages large enough that the energy
dependence of scattering is important.  This is significant from a
practical standpoint-- requiring small voltages limits the
magnitude of the output current and thus the overall scale of the
measurement rate, making the detector more susceptible to
environmental effects.

\subsubsection{General Expression for Noise Correlators}

For completeness, we give explicit expressions for the noise
correlators.  Writing them in terms of energy dependent $N \times
N$ matrix kernels (i.e.~$S_X = \int_{\mu_R}^{\mu_L} d\ve \left[
\textrm{tr } \hat{S}_X(\ve) \right]$) we obtain: \bmlett
\label{Noises}
\begin{eqnarray}
    \hat{S}_I(\ve) & = & \frac{2 e^2}{h} T (1 - T) \\
    \hat{S}_Q(\ve) & = & \frac{e^2 \hbar}{2 \pi} \Bigg(
        \frac{ \left(\partial_\ve T \right)^2 }
            {2 T (1 - T)}
        + 2 T R \left(\phi_U - \phi_V \right)^2
            \nonumber \\
            && + 2
        \left[ \phi_U , \sqrt{T R} \right]
        \left[ \sqrt{T R},\phi_V  \right]
            \nonumber \\
         &&   +
         \left[ \phi_U , T \right]
        \left[ T ,\phi_U \right]
            +
         \left[ \phi_V , T \right]
        \left[ T ,\phi_V \right]
          \Bigg)  \label{MCSQQ}\\
    \hat{\lambda}(\ve) & = &
        -\frac{e^2}{h} \left( \partial_\ve T \right)\\
    \hat{S}_{IQ}(\ve) & = & i \hbar \hat{\lambda}(\ve) +
        \frac{e^2}{\pi}
        \left[
        \sqrt{T R}  \left(
            \phi_U - \phi_V \right) \right]
            \label{MultiMode}
\end{eqnarray}
\emlett%
A similar expression for the charge noise $S_Q$ of a mesoscopic
conductor was first derived by B\"uttiker \cite{EarlyButtiker}.
Unlike the expression for the current noise $S_I$, which can
easily be understood in terms of partition noise, it would seem at
first that there is no simple, heuristic way to interpret the
expression for $S_Q$.  However, if we invoke ideas of information,
each term in Eq.~(\ref{MCSQQ}) acquires a simple meaning.  The
first term represents information associated with the energy
dependence of the transmissions; the second, information
associated with the energy dependence of phase differences; and
the last three terms, information associated with the partitioning
of electrons into different modes.  In general, using
Eqs.~(\ref{Rates}) and (\ref{WeakIA}), we may {\it define} the
charge noise in terms of the accessible information ${\mathcal I}$
in the coupled conductor plus qubit system:
\begin{equation}
    S_{Q} = \lim_{A \ra 0} \lim_{t \ra 0} \frac{\hbar^2}{A^2} \frac{d}{d t}
    {\mathcal I}(t)
\end{equation}
While this last expression may seem purely tautological, it is
clear that the various contributions to Eq.~(\ref{MCSQQ}) for the
charge noise are best understood in terms of information.  Note
that the accessible information ${\mathcal I}$ could be obtained
directly in the present system by calculating the overlap between
the detector states corresponding to the two qubit states.  Such a
calculation would take the form of an orthogonality catastrophe
calculation, similar to that presented in Ref.
\onlinecite{Aleiner}.

\subsection{Local Potential Coupling}

In the remaining part of this paper, we consider a more general
version of the mesoscopic scattering detector, showing that the
main results of the previous section continue to hold. We relax
the condition that the state of the qubit modulates a {\it
uniform} potential in the scattering region, thus allowing for a
wider class of input operators $Q$ than that given in
Eq.~(\ref{QDefn}). In general, we may write:
\begin{eqnarray}
    Q & = &  e \int dE dE' \sum_{\beta, \gamma = L,R}
      \\  \nonumber
     && \left[ a_{\beta n}^{\dagger}(E) W_{\beta n, \gamma m}^{\pd}(E,E')
        a_{\gamma m}^{\pd}(E') \right]
    \label{GQDefn}
\end{eqnarray}
where $W(E,E')$ is a $2N \times 2N$ Hermitian matrix having
dimensions of inverse energy.  The situation considered in the
last section corresponds to choosing $W$ to be $\NN(E,E')$
(Eq.~(\ref{NNDefn})), which at $E=E'$ is just the Wigner-Smith
delay time matrix. By comparing against the current operator $I$
(c.f.~Eq.~(\ref{IDefn})), it is clear that the proportionality
condition of Eq.~(\ref{BigCondition}) necessary for the quantum
limit constrains the diagonal in energy, off-diagonal in lead
index part of the potential matrix $W$:
\begin{eqnarray}
    \forall E \epsilon [\mu_R,\mu_L],
    \hspace{0.2 cm}
    \left[W(E,E)\right]_{R L} = i  \frac{1}{\CC}
    \left[s_{LR}
    \right]^{\dagger}(E) s_{LL}^{\pd} (E)
        \label{GSCondition}
\end{eqnarray}
where $\CC$ is a real constant . We thus see that the required
proportionality between $I$ and $Q$ needed to reach the quantum
limit at zero temperature leaves a large part of the potential
matrix $W$ undetermined (i.e.~terms diagonal in the lead index
and/or off-diagonal in energy).  We now show that by considering a
form for $W$ which is drastically different from $\NN$, one can
make it easier to reach the quantum limit and have a reasonable
gain.  In particular, one can work at small voltages without
necessarily having a vanishing gain.

We specialize the discussion to a case which in many ways is the
opposite of having global potential coupling.  We take the
scattering matrix $s$ to be energy-independent over the energy
interval defined by the voltage, and take $W$ to correspond to a
local potential: $W(E,E') = W$ over the energies of interest. In
this case, the scattering matrix $s$ will have one of two
different energy-independent values depending on the state of the
qubit:
\begin{equation}
    s_{\pm} = s_0 \pm e A \left(  \Delta s \right)
\end{equation}
where $s_0$ is the scattering matrix at zero coupling ($A=0$). The
matrix $W$ may be directly related to the change in the scattering
matrix, $\Delta s$ (see Appendix B for a derivation):
\begin{equation}
    W = i  s_0^{\dagger} \left( \Delta s \right)
    \label{DeltaSEqn}
\end{equation}
Note the similarity to the form of $W$ in the global-potential
coupling case (where $W = \NN$); now, the energy derivative $d s /
d E$ has been replaced by the finite difference $\Delta s \equiv
(s_+ - s_-)/ (2 e A)$.

Turning to the conditions needed for the quantum limit, we find
again that the causality properties of the scattering matrices
$s_{\pm}$ ensure $\lambda'=0$ always.  The remaining
proportionality requirement of Eq.~(\ref{BigCondition}) places
constraints on $s_{\pm}$.  These have an analogous form to
Eqs.~(\ref{MCTCondition}) and (\ref{MCPhaseCondition}), but now
the energy derivative $d / d E$ is replaced by the finite
difference $\Delta$ (i.e.~$\Delta X = (X[s_+] - X[s_-])/(2 e A)$):
\begin{equation}
    \frac{\Delta T}
                {T (1 - T)}  =  \CC \times \hat{1}
                \label{GTCondition}
\end{equation}
\begin{equation}
    \sqrt{T} \widetilde{\phi}_U  \sqrt{R} -
    \sqrt{R} \widetilde{\phi}_V  \sqrt{T}  =  0
    \label{GPhaseCondition}
\end{equation}
where $\widetilde{\phi}_U = -i U^{\dagger} (\Delta U)$,
$\widetilde{\phi}_V = -i V^{\dagger} (\Delta V)$ . Importantly,
{\it the above conditions do not involve any energy averaging}, as
we have taken $s$ and $W$ to be energy independent. Nonetheless,
there still is a non-vanishing gain $\lambda$ determined by both
the voltage and the $\Delta T_j$:
\begin{equation}
    \lambda = \frac{e^2 V}{h} \sum_{j} \Delta T_j
\end{equation}
Thus, using a local coupling between the qubit and the scattering
detector makes it easier to reach the quantum limit and have a
sizeable gain-- one can use voltages small enough that
energy-averaging is not a problem, while still having the qubit
modulate the transmissions.  Note that in the single channel case,
all that is needed for the quantum limit is that the state of the
qubit not change the difference between reflected and transmitted
phases: $\Delta (\phi - \beta) = 0$. Also note the various noise
correlators are given by Eqs.~(\ref{Noises}), with the
substitution $d/dE \ra \Delta$.

\section{Conclusions}

We have developed a general set of conditions which are needed for
a detector in the linear-response regime to reach the quantum
limit of detection.  One needs both a restricted proportionality
between the input and output operators of the detector (c.f.
Eq.~(\ref{BigCondition})), and a causal relation between the
output and input (c.f. Eq.~(\ref{GainCondition})).  Applying the
concept of accessible information to the detector, one sees that
deviations from the quantum limit imply the existence of
``missing" information residing in the detector, information which
is not being utilized.  The general conditions of Eqs.
(\ref{BigCondition}) and (\ref{GainCondition}) ensure the
non-existence of such information.  Applying these concepts to the
mesoscopic scattering detector, we find that these general
conditions place restrictions on the form of the detector's
scattering matrix. These restrictions do not involve symmetry
properties usually considered in mesoscopic systems, but are
rather best understood as following from the requirement of having
no missing information.  In the mesoscopic scattering detector,
missing information may reside in the relative phase between
transmission and reflection, in the energy or mode structure of
the transmission probabilities, or in the partitioning of
scattered electrons between different modes. Surprisingly, we find
that an adiabatic point contact conforms to all the conditions
needed for the quantum limit, even when the voltage is large
enough that many modes are involved in transport, and the energy
dependence of scattering is important.

We thank M. Devoret for useful conversations.  This work was
partially supported by ARDA through the Army Research Office,
grant number DAAD19-02-1-0045, by the NSF under Grants No.
DMR-0084501 and DMR-0196503, and by the W. M. Keck Foundation.

\appendix

\section{Accessible Information}

 In this appendix, we provide a simple proof of Eq.~(\ref{Accessible}) for
 the accessible information $\mathcal I$.  Given the two states $| D_{\ua} \rangle$ and
 $| D_{\da} \rangle$, the goal is to
 maximize the classical mutual information $R$ (defined in Eq.
 (\ref{REqn})) over all possible choices of measurements.  A given choice of
 measurement $Y$ corresponds to a choice of basis;
 the probability distributions $p(y_i|\ua)$ and $p(y_i|\da)$ are
 determined by the elements of the corresponding states in this
 basis.  Treating the $p(y_i|\sigma)$ as independent variables
 restricted to the interval $[0,1]$, and using Lagrange multipliers,
 we minimize $R$ subject to the following constraints:
 \begin{eqnarray}
    \sum_{i=1}^N p(y_i | \sigma) & = & 1 \label{NormCondition}\\
    \sum_{i=1}^N \sqrt{ p(y_i | \ua) p(y_i | \da) } & = &
         | \langle D_{\ua} | D_{\da} \rangle | \equiv
          \cos \alpha \label{IPCondition}
\end{eqnarray}
The second condition in principle need only be an inequality, with
the left-hand side being greater than or equal to the right-hand
side; however, it can be verified that the maximum value of $R$
occurs when it is enforced as an equality.  Also note that without
loss of generality, we can choose the inner product appearing in
Eq.~(\ref{IPCondition}) to be real and positive, as $R$ is
independent of the relative phase between the states $| D_\sigma
\rangle$. Finally, we have assumed to start that these states have
at most $N$ non-zero components in the chosen basis. Variation
with respect to $p(y_i|\ua)$ yields the condition:
\begin{equation}
    \log \frac{p(y_i | \ua)}
        { \bar{p}(y_i) }
        + 2 \lambda_{\ua} + \lambda \sqrt{
            \frac{p(y_i | \da)}
            { p(y_i | \ua) } }  = 0,
\end{equation}
with a similar equation emerging from variation with respect to
$p(y_i | \da)$.  $\lambda$, $\lambda_{\ua}$ and $\lambda_{\da}$,
are Lagrange multipliers; $\bar{p}(y_i) = \left[ p(y_i|\ua) +
p(y_i|\da) \right] / 2 $ is the averaged distribution. Subtracting
the $\ua$ and $\da$ equations yields:
\begin{eqnarray}
    \lambda & = & \frac{ \sqrt{ p(y_i|\da) p(y_i | \ua) } }
        { p(y_i|\ua) - p(y_i|\da) } \log
        \frac{p(y_i|\ua)}{p(y_i|\da)}     \nonumber  \\
        & = &  \frac{ \sqrt{1 - \beta_i^2} }
            {2 \beta_i} \log \frac{1 + \beta_i}{1 - \beta_i}
            \label{BetaEqn}
\end{eqnarray}
where we have defined $\beta_i$ via
\begin{equation}
    \beta_i = \frac{ p(y_i | \ua) - p(y_i | \da)}
        { \bar{p}(y_i)}
\end{equation}
$\beta_i$ may be thought of as the amount of information gained in
a measurement {\it given} that the outcome of the measurement is
$y_i$. Now, Eq.~(\ref{BetaEqn}) must hold for each $\beta_i$
($i=1...N$); moreover, the function on the right-hand side is
symmetric in $\beta_i$ and monotone decreasing for $0 \leq \beta_i
\leq 1$.  It thus follows that for each $i$,
\begin{equation}
    (\beta_i)^2 = \textrm{constant} = \sin^2 \alpha
    \label{BetaEqn2}
\end{equation}
The last equality follows from substitution into
Eq.~(\ref{IPCondition}).  Further substitution into
Eq.~(\ref{REqn}) for $R$ yields the expression in Eq.~(\ref
{Accessible}); note that the averaged distribution $\bar{p}(y_i)$
and the relevant number of basis elements $N$ do not appear in
this expression. One can explicitly check that choosing any of the
$p(y_i|\sigma)$ to be $0$ or $1$ results in a lower value of $R$;
thus, Eq.~(\ref{Accessible}) does indeed correspond to the maximum
value of $R$ and thus, by definition, to the accessible
information $\mathcal I$.  The condition Eq.~(\ref{BetaEqn2})
required to optimize $R$ implies that the amount of information
gained via measurement is the same for each of the measurement
outcomes $y_i$.  Equivalently, each basis element in an optimal
basis has the same information content associated with it. This is
similar to requirements obtained to have the mesoscopic scattering
detector reach the quantum limit; in that case, each channel and
each energy were required to have the same information content
(c.f. Eq.~(\ref{MCTCondition})). Note also that {\it there are
several distinct choices of bases (i.e.~measurement schemes) which
optimize $R$}; this point was not made in Ref.
\onlinecite{Levitin}.  A particularly simple optimal basis can be
constructed for $N=2$. In this basis, the non-zero components of
the states $| D_{\sigma} \rangle$ are given by:
\begin{equation}
    | D_{\ua} \rangle = ( \cos \theta , \sin \theta )
    \hspace{1 cm}
    | D_{\ua} \rangle = ( \sin \theta , \cos \theta )
\end{equation}
where $\theta = \pi/4 + \alpha/2$.  By definition, the state
$(1,0)$ leads to the measurement outcome $y_1$ with perfect
certainty, while the state $(0,1)$ leads to the measurement
outcome $y_2$ with perfect certainty.  In geometric terms, the
optimal basis given here is one in which the angle between the two
states $|D_{\sigma} \rangle$ is bisected by the vector $(1,1)$.

More generally, consider the form of an optimal basis where $N=M$
(i.e. there are $M$ possible outcomes when a measurement is made
on the state $|D_{\ua} \rangle$ or $|D_{\da} \rangle$). Taking $M$
to be even for simplicity, and letting $| j \rangle$ denote the
basis states, a possible optimal basis is one in which:
\begin{eqnarray}
    \langle j | D_{\ua} \rangle & = & \sqrt{
        \frac{1 + (-1)^j \sin \alpha}{M}} \\
    \langle j | D_{\ua} \rangle & = & \sqrt{
        \frac{1 - (-1)^j \sin \alpha}{M}}
\end{eqnarray}
The fact that there are many possible outcomes of a measurement
does not degrade from the optimality of mutual information $R$, as
the information associated with each measurement outcome is the
same.

\section{Derivation of $\Delta s$}

In this Appendix, we provide a brief derivation of
Eq.~(\ref{DeltaSEqn}) which relates the coupling potential matrix
$W$ (c.f.~Eq.~(\ref{GQDefn}) ) to the associated change in the
scattering matrix, $\Delta s$.  The latter quantity determines the
noise correlators and gain of the local-potential coupling version
of the mesoscopic scattering detector.  Our approach is similar to
that used in Ref. \onlinecite{AGB} to relate the scattering matrix
of a quantum dot to its Hamiltonian.

In what follows, we assume (as in Sec. II B) that the potential
matrix $W$ and the zero-coupling scattering matrix $s$ are
independent of energy on the scales of interest.  We start by
writing the system Hamiltonian in terms of the scattering states
of problem at zero coupling, assuming the qubit is frozen in the
$\ua$ state:
\begin{eqnarray}
    H & = &
        \hbar v_F \sum_m \int dk \Bigg[ k\,\
            \psi^{\dag}_m(k) \psi^{\pd}_m(k) + \nonumber \\
        &&
        (A e)  \sum_{m'} \int dk'
        \left(
            \psi^{\dag}_{m'}(k') W_{m' m} \psi^{\pd}_{m}(k)
            \right) \Bigg]
\end{eqnarray}
We have assumed a linear dispersion near the Fermi energy, with
$\hbar k$ and $\hbar k'$ representing the deviation of the
momentum from the Fermi momentum.  We have also neglected the fact
that the effective Fermi velocity is channel dependent ($v_F$
drops out of all final expressions).  The operator
$\psi^{\dagger}_{m}(k)$ creates a scattering state incident in the
lead and transverse mode indexed by $m$.  For definiteness, we
take our leads (both left and right) to be defined only on the
half-line $x < 0$, and to be confined in the $y$ and $z$
directions.  Further, we assume that the scattering region is
situated on $x>0$. We may write the full electron field operator
in terms of the $\psi_m(k)$ operators, using the zero-coupling
scattering matrix $s$.  Writing $\vec{x} = (x,y,z)$, we have:
\begin{eqnarray}
    \Psi(\vec{x}) & = & \sum_m \int \frac{dk}{\sqrt{4\pi}} \psi_m(k) \Bigg[
        e^{i (k_F+k) x} \phi_m(y,z) + \nonumber \\
       &&  \sum_n e^{-i(k_F+k)x} \phi_n(y,z) s_{n m} \Bigg]
       \\
    & = & \frac{1}{\sqrt{2}}\sum_m \psi_m(-x) e^{i k_F x} \phi_m(y,z) +
    \nonumber \\
          && \frac{1}{\sqrt{2}} \sum_{m,n} \psi_m(x) e^{-i k_F x} \phi_n(y,z) s_{n m}
          \label{AuxFieldEqn}
\end{eqnarray}
In the last line, we have introduced the operators $\psi_m(x)$,
which are the Fourier transforms of the scattering state operators
$\psi_m(k)$.  Note again that this expression is only valid for
$x<0$, as the leads are only defined on $x<0$.  We thus see that
for $x<0$, $\psi_m(x)$ describes an {\it outgoing}
(i.e.~left-moving) wave, while $\psi_m(-x)$ describes an {\it
incoming} (i.e.~right-moving) wave.

Next, we may express the system Hamiltonian in terms of the
$\psi_m(x)$ operators.  This in turn leads to an equivalent
single-particle Schr\"odinger equation:
\begin{eqnarray}
    E \psitild_m(E,x)  = &&  \hbar v_F \Big[ i \partial_x  \psitild_m(E,x)
        + \nonumber \\
        && A e \delta(x) \sum_n W_{m n} \psitild_n(E,x) \Big]
\label{Schrodinger}
\end{eqnarray}
Here, $\psitild_m(E,x)$ is a wavefunction which arises when the
field operator $\psi_m(x)$ is expressed in terms of operators
corresponding to the eigenmodes of the full Hamiltonian $H$. Given
the relation of $\psi_m(x)$ to incoming and outgoing waves
(c.f.~Eq.~(\ref{AuxFieldEqn})), we choose the following form for
$\psitild_m(x)$:
\begin{equation}
    \psitild_m(E,x) =
  \begin{cases}
    e^{-i k x} a_{\textrm {in},m} & \text{if $x > 0$}, \\
    e^{-i k x} \sum_n s^{\dag}_{m n} a_{\textrm {out},n} & \text{if $x <
    0$},
  \end{cases}
  \label{ansatz}
\end{equation}
where $E = \hbar v_F k$.  Substituting this form into
Eq.~(\ref{AuxFieldEqn}), we see that the coefficients
$a_{\textrm{in},m}$ and $a_{\textrm{out},m}$ do indeed correspond
(respectively) to the amplitudes of incoming and outgoing waves.

Integrating Eq.~(\ref{Schrodinger}) from $x=0^-$ to $x=0^+$,
interpreting $\psitild(0)$ as $[\psitild(0^+) + \psitild(0^-)
]/2$, and then using Eq.~(\ref{ansatz}), we find the following
relation between the amplitude of incoming and outgoing waves:
\begin{eqnarray}
    a_{\textrm{out},m} & = &
        \sum_{n,n'}  s_{m n}
            \left[
                \frac{ 1 - \frac{i}{2}  A e \widehat{W} }
                { 1 + \frac{i}{2}  A e \widehat{W} }
            \right]_{n n'}
        a_{\textrm{in},n'} \\
        & \equiv & \sum_{n'} \left[s + A e \Delta s
        \right]_{m n'} a_{\textrm{in},n'}
\end{eqnarray}
In the last line, we indicate that this relation defines the new
scattering matrix $s + A e \Delta s$ which includes effects of the
additional potential $W$.  Expanding to lowest order in the
dimensionless potential $A e W$, we find Eq.~(\ref{DeltaSEqn}) as
advertised.


\begin{references}
\bibitem{TwoFactorNote}
    Note that in Ref. \onlinecite{DevoretNature}, the quantum
    limit is written as $\chi < 2$.  This results from using a definition
    of $\Gamma_{\rm meas}$ which is twice as large as the one used
    here.
\bibitem{PureCaveat}
    More generally, there will be two different detector density
    matrices $\rho_{D\ua},\rho_{D\da}$ corresponding to the two
    qubit states; we consider the case of pure states for
    simplicity.
\bibitem{ConstantNote}
    Note that our Eq.~(\ref{TCondition}) has a constant on the right-hand side, while
    the corresponding equation in Ref. \onlinecite{Pilgram} (Eq.~(12)) has an arbitrary
    energy-dependent function on the right-hand side.  This
    difference results from the fact that Ref.
    \onlinecite{Pilgram} considers voltages so small that the
    effects of energy averaging can be neglected.
\bibitem{SymmNote}
    Note that by left-right symmetry, we mean that the {\it
    scalar} potential $V(x,y,z)$ in the scattering region is a
    symmetric function of $x$, where $x$ is the transport
    direction.  One can still have an absence of time-reversal
    symmetry if, e.g., there is a vector potential in the
    scattering region.
\end{references}
\end{document}